\documentclass[aps,prb,showpacs,twocolumn,superscriptaddress]{revtex4-1}
\usepackage{graphicx}
\usepackage{amssymb}

\newcommand{\dgr}{\,$^{\rm o}$}

\begin{document}

\title{Anisotropic magnetism, superconductivity, and the phase diagram of Rb$_{1-x}$Fe$_{2-y}$Se$_2$}

\author{V. Tsurkan}

\affiliation{Experimental Physics 5, Center for Electronic Correlations and Magnetism,
 Institute of Physics, University of Augsburg, D 86159, Augsburg, Germany}

\affiliation{Institute of Applied Physics, Academy of Sciences of Moldova, MD 2028, Chisinau, R. Moldova}

\author{J. Deisenhofer}
\author{A. G\"unther}
\author{H.-A. Krug von Nidda}
\author{S. Widmann}
\author{A. Loidl}

\affiliation{Experimental Physics 5, Center for Electronic Correlations and Magnetism, Institute of Physics,
University of Augsburg, D 86159, Augsburg, Germany}

\date{05.10.2011 [Received: date / Revised version: date ]}

\begin{abstract}

We report the crystal growth and structural, magnetic, conductivity,
and specific heat investigations of Rb$_{1-x}$Fe$_{2-y}$Se$_2$
single crystals with varying stoichiometry prepared by self-flux and
Bridgman methods. The system exhibits a strongly anisotropic
antiferromagnetic behavior below 400~K.  Bulk superconductivity is found in samples with Fe concentrations 1.53 $<2-y<$ 1.6,  whereas for $2-y<$ 1.5
and $2-y>$ 1.6 insulating and semiconducting behavior is observed, respectively. Within the measured range of variation of the Rb concentration (0.6--0.8) no correlation between the Rb content and the lattice parameters of the samples was found. The superconducting samples show the smallest value of the lattice parameter \emph{c} compared to the non-superconducting samples. The sharpest transition to the superconducting state, the highest transition temperature $T_c$ of 32.4~K, and the highest diamagnetic response corresponding to a critical current density $j_c$ of $1.6\times 10^4$~A/cm$^2$ (at 2~K) is found for
compositions close to Rb$_2$Fe$_4$Se$_5$. Upper critical fields $H_{c2}$ of $\sim 250$~kOe for the in-plane and 630~kOe for the
inter-plane configurations are estimated from resistivity studies in
magnetic fields. In the non-superconducting samples with the Fe concentration below 1.45 both specific heat and susceptibility revealed an anomaly at 220~K which is not related to antiferromagnetic or structural transformations. Comparison with the magnetic behavior of
non-superconducting samples provides evidence for the coexistence of
superconductivity and static antiferromagnetic order.
\end{abstract}


\maketitle

\section{Introduction}
Iron-based superconductors\cite{KWH08,RTJ08,TTL08,HLY08} are
currently among the most intensively studied materials. Among
different groups of these superconductors, the iron chalcogenides
have recently attracted particular attention. The undoped iron
selenide, FeSe, exhibts a relatively low critical temperature $T_c$
$\simeq 8$~K at ambient pressure,\cite{HLY08} but it raises to 37~K
under external pressure.\cite{MMT09,MTO09} Earlier attempts to
increase $T_c$ of FeSe using chemical doping resulted in a $T_c$
$\simeq 14$~K by substitution of Se with Te.\cite{FPQ08,YHH08}
Recent reports of $T_c$ $\simeq 30$~K in potassium intercalated FeSe
\cite{JG10} further pushed the research activity in the iron
chalcogenide family. Consequently, successful intercalation of other
alkali metals (Rb and Cs) in FeSe was realized and superconducting
(SC) samples with $T_c$ between 27 and 33~K were
prepared.\cite{KM11,JJY10,CHL10,XGL11,YM11,AFW11,HW11} Further
studies of SC chalcogenides with hypothetical stoichiometry
A$_{0.8}$Fe$_2$Se$_2$ (A = K, Rb, Cs, Tl) revealed significant
differences in their SC properties compared to the related SC
pnictides with a similar structural arrangement.

Intriguing coexistence of superconductivity and static
antiferromagnetic order and a proximity to an insulating state were
suggested.\cite{ZS11,LL11,RHL11,MHF11,WB11} In addition, angle-resolved photoemission spectroscopy (ARPES)
studies \cite{DM11,LZ11,TQ11} showed a different topology of the
Fermi surface in A$_{0.8}$Fe$_2$Se$_2$ compared to other Fe-based
superconductors suggesting a pairing mechanism distinct from
\emph{s}$\pm$ symmetry.\cite{IM11} Despite  numerous reports, the
data on A$_{0.8}$Fe$_2$Se$_2$ materials with regard to their
intrinsic properties is far from being conclusive. For example, the
conducting and magnetic properties of the Rb-based samples reported
by different authors vary significantly indicating a strong dependence on
preparation conditions and impurity content.\cite{CHL10, XGL11, AFW11,HW11,RHL11,MHF11}
Moreover, the correlation between the properties and stoichiometry of
A$_{0.8}$Fe$_2$Se$_2$ has been not fully established yet. Here we present
the results of the structural, magnetic, conductivity, and
thermodynamic characterization of the Rb-Fe-Se system performed on
single crystals grown by two different methods: self-flux and
Bridgman techniques. The selection of the Rb-based system is
motivated by an ease to grow large single crystals with a higher
volume fraction of the SC phase compared to K- and Cs-based Fe
chalcogenides.  The variations of the conducting and magnetic
parameters with the stoichiometry determined by wave-length
dispersive x-ray electron-probe microanalysis (WDS EPMA) are
summarized in a phase diagram of the Rb-Fe-Se system.

\section{Results and discussion}

\subsection{Preparation and composition analysis}
Polycrystalline FeSe synthesized from the high-purity  elements
(99.98 \% Fe and 99.999 \% Se) and 99.75 \% Rb were used as starting
material for the growth of single crystals. Handling of the reaction
mixtures was done in an argon box with residual oxygen and water
content less than 1 ppm. The preparation conditions are given in
Table \ref{tab1}. In the growth runs with starting composition
corresponding to nominal stoichiometry Rb$_{0.8}$Fe$_2$Se$_2$ the
solidified ingots showed significant inhomogeneity. Samples from the
top of the ingots revealed superconducting properties whereas the
bottom of the ingots contained poorly crystallized material with
ferromagnetic behavior. A similar separation was observed for the
growth runs with the starting stoichiometry Rb$_{0.9}$Fe$_2$Se$_2$,
although the samples from the top of the ingot were not
superconducting. In the growth runs with the starting stoichiometry
Rb$_{0.8}$Fe$_{1.6}$Se$_2$, such a strong phase separation was not
observed, but the samples were also non-superconducting.

\begingroup
\squeezetable
\begin{table*}[t]
\caption{Preparation conditions, EPMA, and x-ray data for
Rb$_{1-x}$Fe$_{2-y}$Se$_2$ samples. (Numbers in brackets indicate
standard deviations.)}\label{tab1}
\begin{tabular}{|l|c|c|c|c|c|c|c|c|c|c|}
\hline
Sample & \multicolumn{4}{c|}{Preparation conditions} & \multicolumn{3}{c|}{Concentration}
 & Starting mixture & \multicolumn{2}{c|}{Lattice constant} \\
\cline{2-8} \cline{10-11}
& Method & Soaking & Soaking & Cooling rate & Rb & Fe & Se & & \emph{a, b} & \emph{c} \\
& & temperature & time &  & (1-\emph{x}) & (2-\emph{y}) &  & &
[{\AA}] &[{\AA}] \\ \hline F 266 & self-flux & 1030\dgr\,C & 3 h &
6\dgr\,C/h & 0.653(30) & 1.532(8) & 2.000(35) & Rb+2.5FeSe &
3.9285(3) & 14.6003(18) \\ \hline F 274 & self-flux & 1030\dgr\,C &
3 h & 6\dgr\,C/h & 0.672(15) & 1.430(8) & 2.000(17) & Rb+2FeSe+0.5Se
& 3.9270(9) & 14.6155(49) \\ \hline F 286 & self-flux & 1030\dgr\,C
& 3 h & 6\dgr\,C/h & 0.609(22) & 1.608(9) & 2.000(33) &
Rb+2.5FeSe+0.5Se & 3.9180(11) & 14.6273(61) \\ \hline F 295 &
self-flux & 1030\dgr\,C & 3 h & 6\dgr\,C/h & 0.724(18) & 1.560(10) &
2.000(19) & Rb+2.5FeSe & 3.9310(7) & 14.5914(36) \\ \hline BR 16-1 &
Bridgman & 1070\dgr\,C & 3 h & 3 mm/h & 0.796(28) & 1.596(8) &
2.000(37) & Rb+2.5FeSe & 3.9228(7) & 14.5909(38) \\ \hline BR 16-8 &
&  &  &  & 0.801(28) & 1.602(9) & 2.000(37) &  & 3.9304(6) &
14.6050(28) \\ \hline BR 16-10 &  &  &  & & 0.759(42) & 1.601(9) &
2.000(47) &  &  &  \\ \hline BR 17-1 & Bridgman & 1070\dgr\,C & 3 h
& 3 mm/h & 0.737(26) & 1.439(8) & 2.000(38) & 0.9Rb+2FeSe &
3.9117(5) & 14.6516(36) \\ \hline BR 17-4 &  &  &  & & 0.733(24) &
1.439(9) & 2.000(38) &  &  &  \\ \hline BR 17-5 &  &  &  &  &
0.733(24) & 1.432(15) & 2.000(23) &  &  &  \\ \hline BR 17-10 &  & &
&  & 0.786(17) & 1.459(10) & 2.000(15) &  &  &  \\ \hline BR 18 &
Bridgman & 1070\dgr\,C & 5 h & 3 mm/h & 0.659(32) & 1.537(9) &
2.000(36) & Rb+2.5FeSe & 3.9282(4) & 14.5899(23) \\  \hline BR 19 &
Bridgman & 1070\dgr\,C  & 5 h & 3 mm/h & 0.698(17) & 1.497(11) &
2.000(25) & Rb+2FeSe+0.5Se & 3.9379(1) & 14.5983(7)\\  \hline BR 22
& Bridgman & 1070\dgr\,C  & 5 h & 3 mm/h & 0.644(31) & 1.620(12) &
2.000(35) & Rb+2.5FeSe+0.5Se & 3.9021(7) & 14.6484(58)\\  \hline BR
26 & Bridgman & 1070\dgr\,C & 5 h & 3 mm/h & 0.740(36) & 1.600(6) &
2.000(25) & Rb+2.5FeSe & 3.9190(9) & 14.5515(57)  \\ \hline BR 28 &
Bridgman & 1070\dgr\,C & 5 h & 3 mm/h & 0.685(21) & 1.588(9) &
2.000(40) & Rb+2.5FeSe & 3.9251(6) & 14.5932(36)  \\ \hline
\end{tabular}
\end{table*}
\endgroup

The deviation from the nominal stoichiometry is a general problem
for the intercalated iron chalcogenides. In most earlier reports, the
composition of samples was determined by semi-quantitative EDX
analysis or more precise ICP analysis. Both methods of compositional
analysis revealed a different stoichiometry compared to the nominal
Rb$_{0.8}$Fe$_2$Se$_2$.\cite{CHL10, XGL11, AFW11} Various ranges of iron
stoichiometry for the superconducting samples were reported, but the accuracy
of these methods appears not to be sufficient to detect subtle changes in the
composition. Therefore, the composition of the grown samples was
determined by WDS EPMA that allows to reach an accuracy of 0.5~\%
for Fe and 1~\% for Se. The concentration of the elements in the
samples was measured using a Cameca SX50 analyzer on freshly cleaved
samples minimally exposed to air ($\sim 1$ min) to prevent oxidation.
The EPMA data are also given in Table \ref{tab1}. They
represent the values averaged over multiple (at least 10) measured
spots with an area of $80\times60$ $\mu $m$^2$. The concentrations of Rb
and Fe elements are normalized assuming a Se concentration of 2 per formula unit. The
EPMA analysis revealed notable deviations in the composition of the
samples from the starting stoichiometry for all prepared batches.
The concentration of Fe in the superconducting samples from
different batches was found to vary between 1.53 and 1.60, while the
variations of the Rb concentration were between 0.65 and 0.8. The Rb
deficiency is probably caused by a strong interaction with quartz
glass at high temperatures. The stoichiometry of the samples
exhibiting the highest SC parameters (from batches BR~16 and BR~26
described below) was close to Rb$_{0.8}$Fe$_{1.6}$Se$_2$ (or
Rb$_{2}$Fe$_{4}$Se$_5$ in other notation). The concentration of Fe
in the samples prepared from exactly the same nominal mixture
Rb$_{2}$Fe$_{4}$Se$_5$ was, however, less than $\sim1.5$ and the
samples were not superconducting. Another group of
non-superconducting samples was also found, with the Fe
concentration being higher than $\sim1.6$. The variations of the Fe
and Se concentrations on the individual samples were less than 1 and
2~\%, respectively, indicating their high homogeneity. The maximal
variations of the Rb concentrations were of about 5~\% which may indicate an inhomogeneous distribution of Rb ions. However, we do not exclude that this can be also an artifact due to observed sensitivity of the counting rate to a slight misalignment of the sample due to remaining fragments of the removed layers on the surface after cleavage and strong sensitivity of Rb to oxidation.

\subsection{X-ray diffraction and structure}
Figures 1 and 2 demonstrate the room temperature x-ray powder
diffraction patterns obtained on crushed single crystals using a
STOE Stadi P diffractometer with Cu K$_\alpha$ radiation, $\lambda =
1.54056$~{\AA} for superconducting and non-superconducting samples,
respectively. The lattice parameters of the samples were obtained
from the Rietveld refinement using the FULLPROF SUITE \cite{RC93}
assuming tetragonal symmetry I4/mmm. They are presented in Table
\ref{tab1}. The refinement was done using the site occupations for
constituent elements as determined by EPMA.

\begin{figure}[t]
\centering
\includegraphics[angle=0,width=0.45\textwidth]{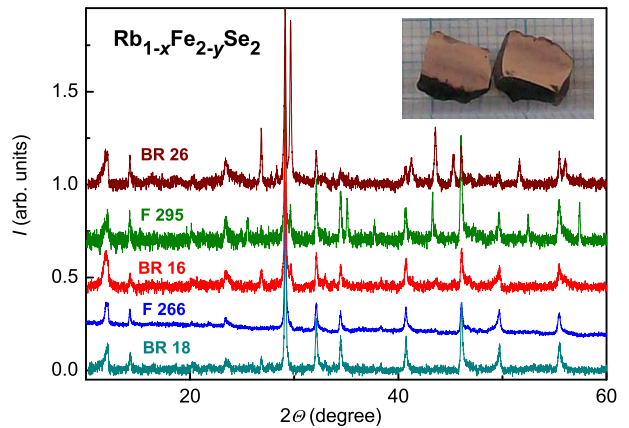}
 \caption{(color online) X-ray diffraction patterns for different superconducting Rb$_{1-x}$Fe$_{2-y}$Se$_2$ samples.
 Inset: image of the single crystals BR~26 subtracted from the ingot. Sample labels BR and F stand for those grown by Bridgman and flux methods, respectively.}
\end{figure}

\begin{figure}[t]
\centering
\includegraphics[angle=0,width=0.45\textwidth]{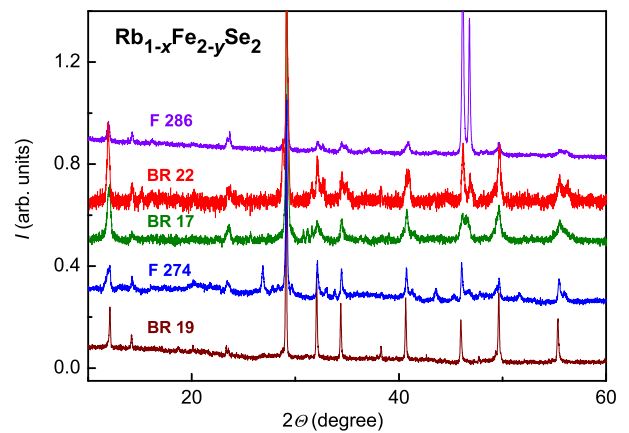}
 \caption{(color online) X-ray diffraction patterns for different non-superconducting
 Rb$_{1-x}$Fe$_{2-y}$Se$_2$ samples.}
\end{figure}

It is important to note that non-superconducting samples from the batch BR~19 with Fe concentration $\simeq 1.5$ show the narrowest diffraction lines. X-ray spectra of the non-superconducting samples from other batches show additional features which are probably related to the difference in their vacancy structure. For example, samples with the Fe concentration below 1.45 exhibit much broader peaks indicating inhomogeneity and stress. Samples with the Fe
concentration above 1.61 exhibit splitting of peaks which can be attributed to orthorhombic distortions. We also noticed that superconducting samples exhibit much broader x-ray diffraction peaks and additional reflexes beside the main peaks. In the superconducting samples from batches F~295, BR~16 and
BR~26, the additional reflexes are of a considerable magnitude. Additional peaks, although of smaller magnitudes, were also observed
in the single crystal x-ray diffraction spectra in a number of SC
A$_{0.8}$Fe$_{1-x}$Se$_2$ compounds but at lower angles compared to
the main peaks.\cite{JJY10,XGL11,HW11,ZG11} These additional peaks
were attributed either to inhomogeneous distribution of the alkali
ions or to a modulated structure resulting from ordering of iron
vacancies. The absence of such peaks in the non-superconducting samples probably exclude any significant inhomogeneous
distribution of the Rb ions in our samples.
Recent single crystal x-ray and neutron diffraction studies of a number of
A$_{0.8}$Fe$_2$Se$_2$ compounds identified their crystal
structure within a tetragonal I4/m space group
symmetry.\cite{WB11,PZA11,VYP11,FYE11}  The
x-ray patterns of our samples were also refined within the I4/m symmetry. For example, for the SC sample BR~16 the obtained value of the lattice constant
\emph{a}=8.7977~{\AA} agrees well with $\sqrt5\emph{a}$ of the
supercell. We must note  that  x-ray pattern only of the non-superconducting sample BR~19 with the sharpest diffraction peaks is fully compatible with the I4/m space group symmetry. However, within the I4/m symmetry it was not possible to refine additional peaks observed for the superconducting samples BR~16 and BR~26, suggesting a second crystallographic phase coexisting with the main phase. The coexistence of different crystallographic phases,
was earlier reported for non-superconducting K$_{1-x}$Fe$_{2-y}$Se$_2$.\cite{WBao11} However, the
orthorhombic symmetry \emph{Pmna} of the additional phase in Ref.~\onlinecite{WBao11} is incompatible with the symmetry of the
second crystallographic phase observed in our SC samples. The phase separation in intercalated iron chalcogenides is now actively investigated. Recent transmission electron microscopy,\cite{ZWa11} x-ray diffraction,\cite{ARi11}  ARPES,\cite{FCh11} optical spectroscopy,\cite{RHY11} and scanning tunneling microscopy studies \cite{WLi11, PCa11} provided evidence for the phase separation in K-intercalated iron superconductors. Very recently, inelastic neutron scattering,\cite{JTP11} optical spectroscopy,\cite{ACh11} and M\"{o}ssbauer studies \cite{VKs11} of the superconducting Rb$_{0.8}$Fe$_{1.6}$Se$_2$ samples (batch BR~16) confirmed the  phase separation scenario also for the Rb-based superconductors. However, the relevance to phase separation of the second phase observed in the x-ray spectra of the superconducting samples is still an open question and additional studies are necessary to clarify its origin.

Since the quality of the refinement of the x-ray powder-diffraction data within the I4/m symmetry was lower compared to the I4/mmm symmetry
due to reduced statistics and a larger number of refined parameters, we present the data of the refinement within the I4/mmm symmetry. The analysis of the obtained data shows only slight variations of the lattice parameter
$\emph{a}$ with the Fe concentration in the range 1.53--1.6 and a
decreasing tendency below 1.5 and above 1.6. An opposite trend is
found for the variation of the \emph{c}-parameter. The largest value
of the lattice parameter $\emph{a}$ and the largest value of the
unit cell volume are observed for the non-superconducting sample BR~19 with Fe concentration $\simeq1.5$. With exception of this sample,
the values of the \emph{c}-parameter are lower for the
superconducting samples, compared to the non-superconducting ones.
Surprisingly, within the measured range of variation of the Rb concentration in the samples (0.6-0.8) no correlation between the Rb concentration
and the value of the \emph{c}-parameter was found, although one
would intuitively expect an increase of the \emph{c}-parameter with
increasing Rb content. This probably can be attributed to a
difference in the defect structure of the samples, which requires a
more comprehensive study. To solve remaining issues concerning the
defect structure, vacancy ordering, and the symmetry of the
additional crystallographic phase, single crystal x-ray
investigations are desirable.

Finally, we are also documenting the results of the microstructural
study. We observed that samples with the Fe concentration in the
range of 1.53--1.6 have very flat surfaces without visible defects
or precipitates. They can be easily cleaved along the
\emph{ab}-plane reflecting the two-dimensional structure of the
crystals. The samples with the Fe content below 1.45 are also rather
easily cleaved; their surface contains a large number of dot-like
defects. The microstructure of the samples with the Fe concentration
above 1.6 is considerably different compared to that of the samples
with a lower Fe content. They are mostly three-dimensional and
therefore difficult to cleave along the \emph{ab}-plane.

\subsection{Magnetic susceptibility}
Figures 3 and 4, respectively, present the magnetic susceptibility
versus temperature for superconducting and non-superconducting
samples from different growth runs measured in a magnetic field
\emph{H}=10~kOe applied parallel to the \emph{c}-axis.  The
measurements were performed on cooling using a SQUID magnetometer
MPMS~5 (Quantum Design). On lowering temperature from 400~K to 33~K
the susceptibility for all SC samples continuously decreases. The SC
samples prepared by self-flux from the starting composition
Rb$_{0.8}$Fe$_2$Se$_2$ exhibit a higher susceptibility than those
prepared by Bridgman method, which probably can be  attributed to
the presence of a small amount of non-reacted iron, e.g., of
$\sim0.3 \%$ estimated from the magnetic moment for the sample F 266
measured at 40~K that exhibits saturation at low fields. In
contrast, the magnetization of the SC samples grown by Bridgman
method shows non-saturated  behavior (see below) and the amount of
the non-reacted iron in these samples appears to be significantly
reduced. In the SC samples a clear downturn of $\chi$ at the
transition temperature at around 30~K is evidenced, while $\chi$ for
the non-superconducting samples shows a small upturn towards lower
temperatures. It is important to note the very close values and
essentially similar temperature dependence of the susceptibility at
temperatures above 33~K both for the SC samples from batch BR~16 and
non SC samples from batch BR~19 despite a clear difference in their
chemical composition. The susceptibility of the exceptional sample BR~17
(with the strongly reduced Fe concentration below 1.46) shows a
nonmonotonic temperature dependence with a broad
maximum at around $T^*$=233~K (for \emph{H}$\|$\emph{c}), which is also
evidenced in the specific heat (see below).

In Fig.~5, the temperature dependences of the magnetic
susceptibility are shown for two selected SC and non SC samples (BR~16
and BR~19) measured in a magnetic field \emph{H}=10~kOe applied
parallel and perpendicular to the \emph{c}-axis. The susceptibility
$\chi_\bot$ for the perpendicular configuration
(\emph{H}$\bot$\emph{c}) is by a factor of five higher than for
\emph{H}$\|$\emph{c}, and shows a weak temperature dependence unlike
the significant decrease of $\chi_\|$ on decreasing temperature.
Such behavior of the $\chi_\bot$ and $\chi_\|$ is characteristic for
an anisotropic antiferromagnet, with the \emph{c}-axis being the
direction of alignment of the spins. A similar anisotropy of the
magnetic properties has been recently reported for insulating
TlFe$_{1.6}$Se$_2$ \cite{BCS11} with a N\'{e}el temperature of 430~K.
The anisotropic antiferromagnetism appears to be a distinct
feature of the whole family of the intercalated iron chalcogenides.

In order to determine the magnetic ordering temperature we measured
the susceptibility of the sample BR~16 at temperatures up to 650~K
using a heating option of the SQUID. The measurements (not shown)
revealed a poorly pronounced maximum in the susceptibility at a
temperature of 527($\pm1$)~K. At the same time, a strongly
irreversible behavior of $\chi(T)$ on successive cooling and warming
cycles was detected. Subsequent low-temperature measurements of the
sample after the high-temperature cycling showed a strongly reduced
amount of the SC phase indicating a partial degradation of the
sample. Therefore it is unclear whether 527~K represents a true
N\'{e}el temperature for this sample. Thus, to get an independent
estimate of the N\'{e}el temperature we used an extrapolation of the
$\chi(T)$ data for the two measured configurations assuming that at
$T_N$ they should merge. The data on $\chi_\bot$ and $\chi_\|$ above
100~K were fitted by polynomials of second order. From the
extrapolation of $\chi_\bot$ and $\chi_\|$  a N\'{e}el temperature
of about 620 ($\pm 5$)~K was estimated for SC sample BR~16.

\begin{figure}[t]
\includegraphics[angle=0,width=0.45\textwidth]{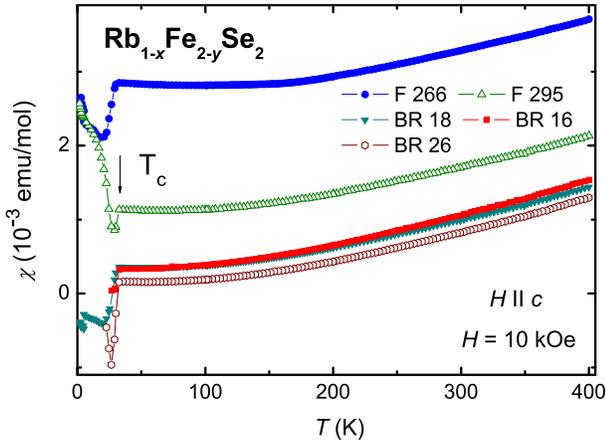}
 \caption{(color online) Temperature dependences of the field-cooled susceptibility
 for superconducting Rb$_{1-x}$Fe$_{2-y}$Se$_2$ samples from different batches
 measured in a field of 10~kOe applied along the \emph{c}-axis. The arrow indicates
 the SC transition temperature. }
\end{figure}

\begin{figure}[t]
\includegraphics[angle=0,width=0.45\textwidth]{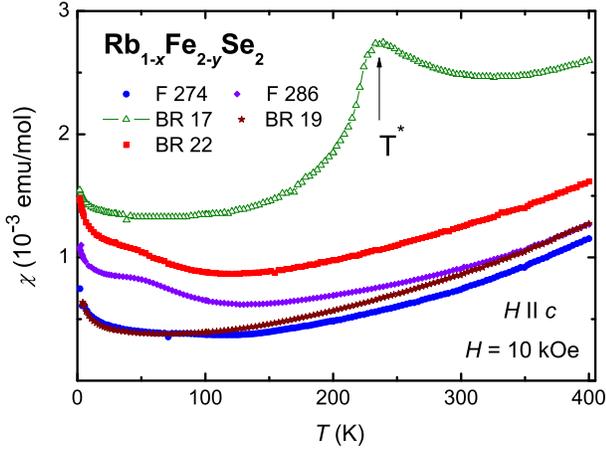}
 \caption{(color online) Temperature dependences of the field-cooled susceptibility
 for non-superconducting Rb$_{1-x}$Fe$_{2-y}$Se$_2$ samples from different batches
 measured in a field of 10~kOe applied along the \emph{c}-axis. The arrow marks the
 temperature of the maximum of the susceptibility $T^*$ for sample BR~17. }
\end{figure}

\begin{figure}[t]
\includegraphics[angle=0,width=0.45\textwidth]{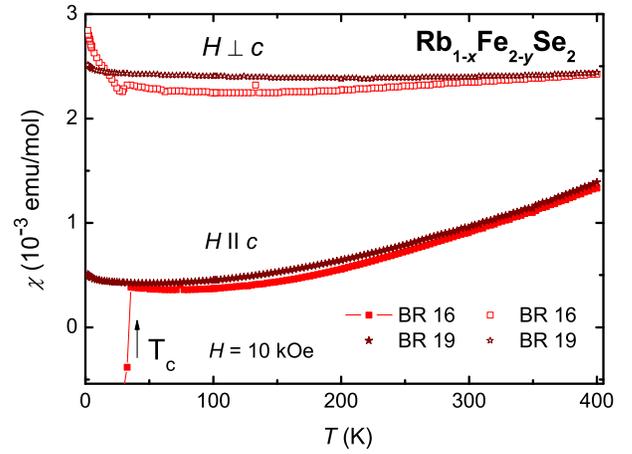}
 \caption{(color online) Temperature dependences of  the field-cooled susceptibility
 for two SC (BR 16) and non SC (BR 19) Rb$_{1-x}$Fe$_{2-y}$Se$_2$ samples
 measured in a field of 10~kOe applied parallel and perpendicular to the \emph{c}-axis. }
\end{figure}

In Figs.~6  and 7, respectively, the magnetization curves
\emph{M}=f(\emph{H}) for the superconducting sample BR~16  and
non-superconducting sample BR~19, measured at different temperatures
are presented. The data are shown for the magnetic field applied
parallel and perpendicular to the \emph{c}-axis. In the SC sample,
\emph{M}$_\|$ shows a small non-linearity at low fields and a linear
increase at high fields. \emph{M}$_\|$ increases significantly with
 temperature from 40 to 400~K. At the same time,
\emph{M}$_\bot$ being considerably higher than \emph{M}$_\|$, shows
a weak temperature dependence in the temperature range 40--400~K in
sharp contrast to \emph{M}$_\|$. It is worth noting here that the
magnetization of the non SC sample exhibits a field and temperature
dependence for both field configurations very similar to that
observed for the SC sample. Importantly, the magnetization of the
non SC sample is not changing in the range from 40 to 2~K as
demonstrated by the coinciding data for 2 and 40~K in Fig.~7. A
close similarity of the magnetization behavior observed  for both
the SC and non SC samples suggests that the static magnetic order
coexists with the superconducting state.

\begin{figure}[t]
\includegraphics[angle=0,width=0.45\textwidth]{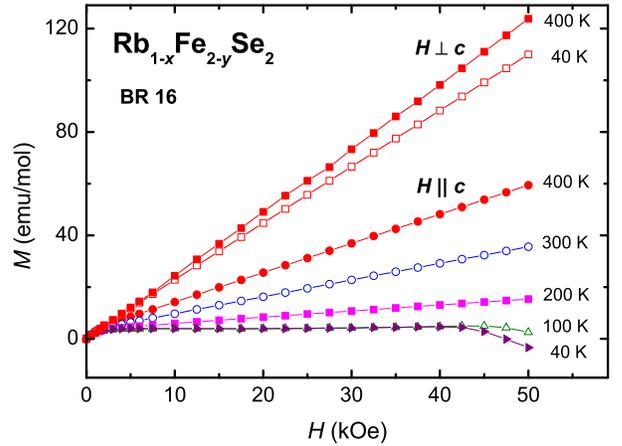}
\caption{(color online) Magnetization curves for the SC sample BR~16
for different temperatures  measured with the magnetic field applied
along the \emph{c}-axis and within the \emph{ab}-plane.}
\end{figure}

\begin{figure}[t]
\includegraphics[angle=0,width=0.45\textwidth]{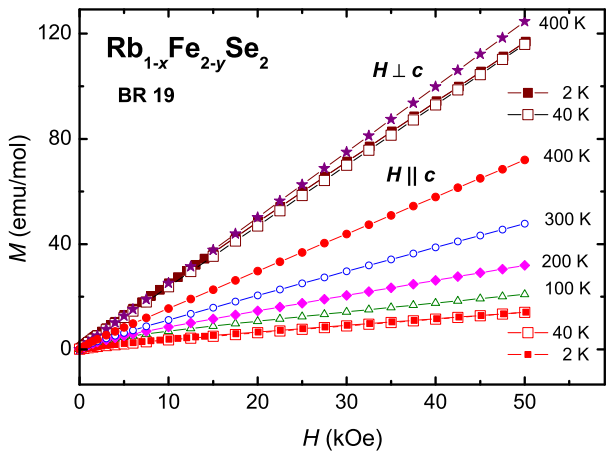}
\caption{(color online) Magnetization curves for the non SC sample
BR~19 measured  at different temperatures  with the magnetic field
applied along the \emph{c}-axis and within the \emph{ab}-plane. Data
for 2~K (full squares) and 40~K (open squares) are almost coinciding. }
\end{figure}

Figure~8 shows the temperature dependences of the zero-field cooled
(ZFC) and field-cooled (FC) susceptibility $\chi$ for
superconducting samples from different batches measured in a field
\emph{H}=10~Oe applied parallel to the \emph{c}-axis. The sharpest
transition ($\Delta{T_c}$ = 1.5~K, estimated from FC data) and the highest temperature of
the onset of superconductivity, $T_c^{on}$= 32.4~K was found for the
samples from batches F~295, BR~16, and BR~26. Although the EPMA data
presented above indicate some notable variations in the Fe and Rb
concentrations for the samples from these batches, no significant
broadening of the SC transition or variation of $T_c^{on}$ was
detected. At the same time, for the SC samples from the other
batches (for example, F~266 and BR~18), the value of $T_c^{on}$ was
by $\sim$2~K lower and the variation of $\chi_{ZFC}$ indicates a broader transition compare to the samples from batches F~295, BR~16, and BR~26. The reduction of the transition temperature and broadening of the transition for these samples probably can be attributed to disorder effects caused, for example, by larger deviations of the Rb and, especially, of the Fe concentrations from the "optimal" values of 0.8 and 1.6, respectively.

These "optimal" values of Rb and Fe concentrations correspond to a formal oxidation state  for the Fe ions of +2.0  as found in other superconducting Fe chalcogenides. Within the experimentally determined ratio of Rb to Fe concentration (see Table 1) we however did not found a change from the hole to electron doping with the variation of the stoichiometry. Indeed, for the samples with the lowest Fe concentration (below 1.5, e.g., BR17) the calculated valence of Fe is +2.27. For the samples with high Fe content (above 1.61, e.g., BR22) the valence of Fe is +2.07. Measurements of the thermoelectric power are necessary to clarify the role of doping on the properties of the Rb-based iron chalcogenides, as it was shown for the related  K-intercalated compounds.\cite{KWa11, YJY11}

It is important to note that the values of $\chi_{FC}$ (Meissner effect) at 2~K
for the SC samples are at least three order of value lower than the values of $\chi_{ZFC}$ (shielding effect), probably due to strong pinning.
At the same time,  the values of $4 \pi \chi$ at 2~K
calculated from the $\chi_{ZFC}$ data are far
above unity pointing to the dominance of demagnetizing effects. The
measurements of $\chi_{ZFC}$ for the in-plane configuration (with
\emph{H} perpendicular to the \emph{c}-axis) with the negligible
demagnetizing factor for the samples from batches F~295, BR~16, and
BR~26 gave a value of $4 \pi \chi$ close to unity that may indicate
a large volume fraction of the SC phase which yields a perfect 100~\% shielding.

\begin{figure}[t]
\includegraphics[angle=0,width=0.45\textwidth]{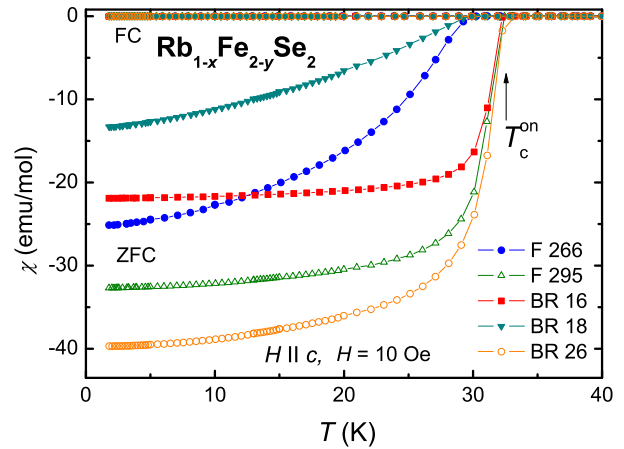}
 \caption{(color online) Temperature dependences of ZFC and FC susceptibilities
 for different superconducting Rb$_{1-x}$Fe$_{2-y}$Se$_2$ samples measured in
 a field of 10~Oe applied along the \emph{c}-axis. The arrow indicates the
 temperature of the onset of the superconducting transition $T_c^{on} = 32.4$~K for sample BR~16.}
\end{figure}

Figure 9 presents the magnetization hysteresis loops for SC samples
measured at 2~K with the magnetic field \emph{H} applied along the
\emph{c}-axis. The hysteresis loop for this configuration is 
symmetric with respect to the origin (0.0). The diamagnetic response for samples from different
batches differs by a factor of 20. The reason of such a strong
difference is unclear. It can be caused by variations of the
stoichiometry detected by EPMA (see Table \ref{tab1}) or by
changes in the preparation conditions, such as oxygen contamination.
The phase separation can also be critical for obtaining high SC
hysteresis width. Indeed, the highest diamagnetic response was found in
the samples from batches F~295, BR~16 and BR~26 in which x-ray
diffraction revealed strong reflexes of the additional crystallographic phase.
Additional pinning centers created by this additional crystallographic phase can enhance the diamagnetic response.

The critical current density $\emph{j}_c$ at 2~K estimated from the
hysteresis data for the samples with the highest SC parameters using
the Bean model for hard superconductors \cite{B62,B64} reaches a
value of $1.6 \times 10^4$~A/cm$^2$. It is by a factor of five lower
than the intrinsic value of $j_c$ reported for FeSe substituted with
Te.\cite{VT10, AGu11}

In Fig.~10 the hysteresis loop for the SC sample BR~16 measured at
2~K with the magnetic field applied perpendicular to the
\emph{c}-axis is plotted. The maximum values of magnetization for
\emph{H} $\bot$ \emph{c} and \emph{H} $\|$ \emph{c} correlate with
the difference in the demagnetization factors for these two
configurations. The hysteresis loop for the configuration \emph{H}
$\|$ \emph{c} reveals two contributions, one
superconducting, and the other linear in magnetic field. For
comparison, in Fig. 10 the magnetization curve
\emph{M}$_\|$=f(\emph{H}) for the same sample measured at 40~K, just
above the SC transition, is shown. It agrees well with the linear
contribution that can be derived from the data at 2~K. This again
points out that the magnetic correlations associated with the static
antiferromagnetic order coexist with the SC state.

\begin{figure}[t]
\includegraphics[angle=0,width=0.45\textwidth]{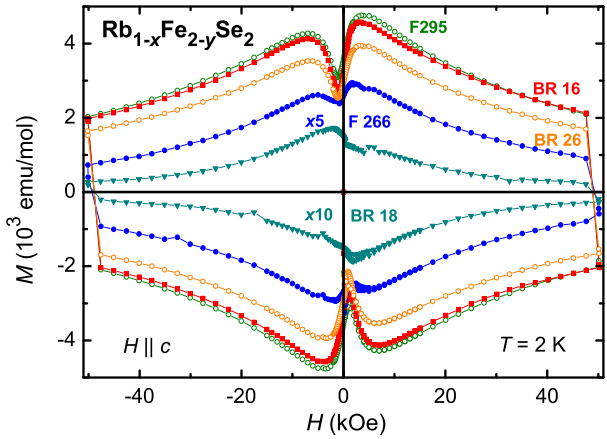}
 \caption{(color online) Hysteresis loops measured at 2~K with magnetic field
 applied along the \emph{c}-axis for different SC samples Rb$_{1-x}$Fe$_{2-y}$Se$_2$.
 For clarity, the data for samples F~266 and BR~18 are magnified by factors of 5 and 10, respectively.
 }
\end{figure}

\begin{figure}[t]
\includegraphics[angle=0,width=0.45\textwidth]{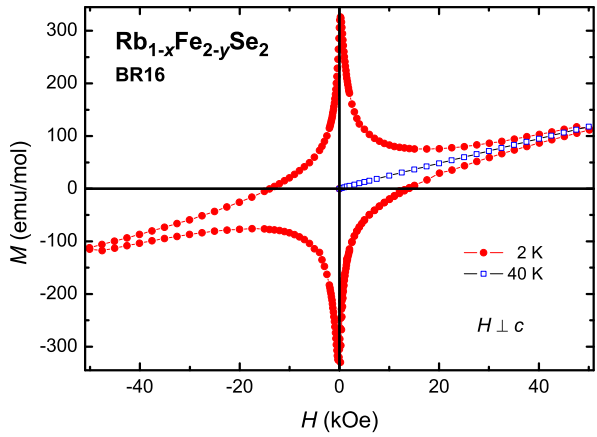}
\caption{(color online) Hysteresis loop for SC sample BR~16 at 2~K
(solid circles) measured with the magnetic field applied perpendicular
to the \emph{c}-axis, together with the magnetization curve at 40~K
(open squares).
 }
\end{figure}

\subsection{Resistivity}
Figures 11(a) and (b) demonstrate the temperature dependences of the
resistivity for the SC and non SC samples, respectively. The
resistivity measurements were done on rectangular samples by a
four-point method in the temperature range 2--300~K using the
resistivity measurement option of the Quantum Design physical
properties measurement system (PPMS) in magnetic fields up to
90~kOe. Both, the in-plane resistivity, $\rho_{ab}$, with current \emph{j} in the \emph{ab}-plane,
and the inter-plane resistivity, $\rho_{c}$, with current \emph{j} parallel to the \emph{c}-axis were measured. The magnetic field was always perpendicular to the current direction, so only the transverse configuration was studied. The contacts were made with a conductive silver paint. The error
in the absolute value of the measured resistivity was about 20~\% due to finite dimensions of the potential contacts.
The in-plane resistivity $\rho_{ab}$ of the SC sample
prepared by self-flux method (F~266) is by approximately one order
of magnitude larger than that for samples grown by Bridgman method which can be attributed to higher impurity content of the self-flux samples.
The inter-plane resistivity $\rho_{c}$ of a typical
sample grown by Bridgman method with high SC parameters (BR~16) is by a factor of two higher
than the in-plane resistivity. The resistivity curves of all superconducting
samples show a broad hump at a temperature $T_{max}$ that for
different samples varies in the range of 190--215~K with a
semiconductor-like temperature dependence above and a metal-like
dependence below $T_{max}$. We note that the superconducting samples from batch
BR~16 have the smallest value of the resistivity compared to other
batches. It is also the smallest for the superconducting
Rb$_{1-x}$Fe$_{2-y}$Se$_2$ crystals reported so
far.\cite{CHL10,AFW11} Together with the larger resistivity ratio
$\rho(T_{max})/\rho(T_c)$ = 37 this may indicate a higher purity of
these superconducting samples.

\begin{figure}[htb]
\includegraphics[angle=0,width=0.45\textwidth]{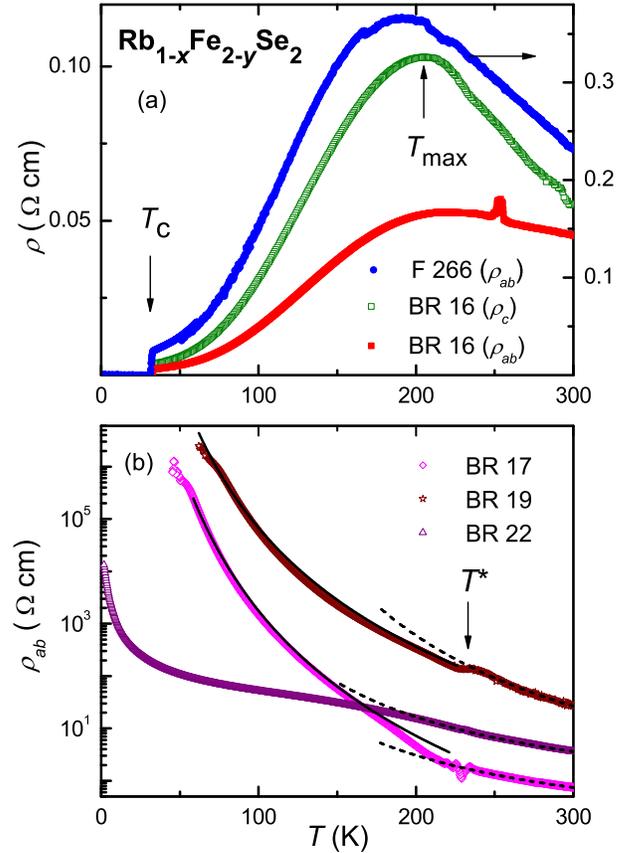}
\caption{(color online) (a) Temperature dependences of the in-plane resistivity $\rho_{ab}$ and inter-plane resistivity $\rho_{c}$ for the SC sample BR~16 (left scale) and for the in-plane resistivity for the SC sample F~266 (right scale) measured on cooling in zero external magnetic field. (b) Temperature
dependences of the in-plane resistivity for the non SC samples.  Dashed curves mark the Arrhenius fits at high temperatures, solid lines - the MVRH fits al lower temperatures as described in the text.}
\end{figure}

The resistivity of the non-superconducting samples is by several orders higher
than that of the SC samples. It shows a steep increase on decreasing
temperature and a small but clearly discernible anomaly at $T^*$=233~K.
In the range 300--240~K the temperature dependence of the
resistivity for the non SC samples can be described by Arrhenius law $\rho=\rho_0\exp(\Delta E_a/kT)$  with an activation
energy $\Delta E_a$ of 0.16, 0.079 and 0.073~eV for the samples
BR~19, BR~22, and BR~17, respectively. At temperatures below 240~K
down to 80~K the resistivity of the non SC samples is not thermally
activated but, instead, can be reasonably described within a Mott
variable-range hopping (MVRH) model by an expression
$\rho=\rho_0\exp(T_0/T)^{1/4}$. The resistivity of the sample
BR~22 with the Fe concentration above 1.62 is also showing
semiconductor-like behavior with decreasing temperature, but its value is by
several orders of magnitude lower than that of the samples with the
Fe concentration below 1.5. Summarizing the observed
resistivity behavior of the non-superconducting samples we notice
rather low values of the activation energies which are more typical
for heavily doped semiconductors or Mott insulators than for
intrinsic band insulators. Additional studies including Hall effect
are necessary to clarify the actual transport mechanisms in the
non-superconducting Rb$_{1-x}$Fe$_{2-y}$Se$_2$ crystals.

Figures 12(a) and (b), respectively, illustrate the effect of a transverse
magnetic field on the in-plane $\rho_{ab}$ and inter-plane
$\rho_{c}$ resistivities in the transition region for the SC sample BR~16. The
measurements were done on warming after cooling in zero field. In
zero field the transition temperature, determined at the level of
the 90~\% drop of the normal state resistivity $\rho_n$, coincides
with the $T_c^{on}$ (within 0.1~K) determined from the
susceptibility measurements. This again indicates the high quality
of the superconducting samples. When increasing the magnetic field, the resistivity
curves are displaced to lower temperatures. The inset in Fig.~12(b)
shows the temperature dependence of the upper critical field
$H_{c2}(T)$ estimated using the criterion of the 90~\% drop of the
normal state resistivity. On approaching $T_c$, the slope of the
$H_{c2}(T)$ curve becomes smaller compared to that at lower
temperatures. The estimation of the upper critical field $H_{c2}(0)$
for $T=0$~K made within the Werthamer-Helfand-Hohenberg model
\cite{WHH66} using the relation $H_{c2}(0)= - 0.69 T_c
(dH_{c2}(T)/dT)\mid_{T_c}$ gave a value of 250~kOe for the in-plane
and 630~kOe for the inter-plane configurations. They are by a factor
of two lower than those for FeSe substituted with Te \cite{VT10, AGu11} and
are the lowest reported so far for SC Rb$_{1-x}$Fe$_{2-y}$Se$_2$
samples.\cite{CHL10,AFW11}
\begin{figure}[htb]
\includegraphics[angle=0,width=0.4\textwidth]{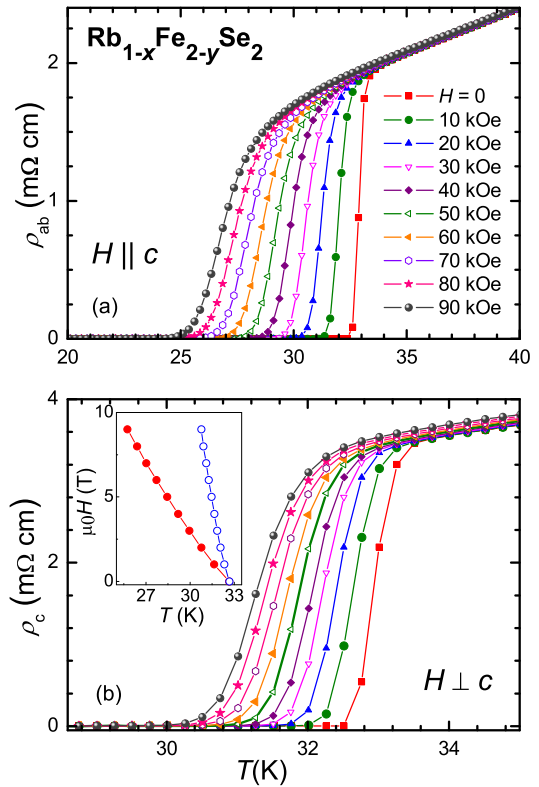}
\caption{(color online) (a) Temperature dependences of the in-plane
resistivity $\rho_{ab}$ for the sample BR~16 measured in various magnetic fields
applied parallel to the \emph{c}-axis. (b) Temperature dependences
of the inter-plane resistivity $\rho_{c}$ for the sample BR~16 measured in
various magnetic fields applied perpendicular to the \emph{c}-axis.
Inset: temperature dependence of the upper critical field for
inter-plane (closed symbols) and in-plane (open symbols)
configurations as described in the text.}
\end{figure}

\subsection{Specific heat}

Figure 13 presents the temperature dependences of the specific heat
\emph{C} for selected superconducting and non-superconducting
samples. The heat capacity was measured by a relaxation method using
the PPMS in the temperature range 1.8--300~K and in magnetic
fields up to 90~kOe. The magnetic field was applied parallel to the
\emph{c}-axis of the samples. The specific heat for different
samples exhibits very similar behavior and values in the measured
temperature range. The specific heat of the non SC sample BR~17
shows a sharp peak at around 220~K. Its position is lower than that
of the peak in the susceptibility at 233~K for the same field
configuration. Interestingly, the susceptibility of this sample for
the field configuration perpendicular to the \emph{c}-axis exhibits
an anomaly exactly at the same temperature of 220~K as observed in
the specific heat. The application of the magnetic field of 50~kOe
does not have any influence on the temperature position of the peak
in the specific heat at 220~K indicating that it is not related to
conventional AFM or FM transitions. Temperature dependent x-ray
measurements in the range 300--100 K did not find any evidence of a
structural transformation either. Preliminary M\"{o}ssbauer studies of
the samples from this batch revealed new lines below $T^*$ typical
for trivalent Fe ions.\cite{VK11} The observed anisotropy in the value
of $T^*$ for \emph{H}$\|$\emph{c}-axis and \emph{H}$\perp$\emph{c}-axis
suggests a very peculiar type of this transition which needs further investigation.

\begin{figure}[htb]
\includegraphics[angle=0,width=0.45\textwidth]{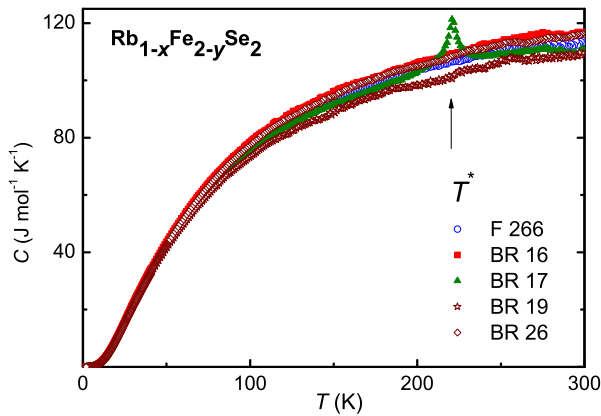}
 \caption{(color online) Temperature dependences of the specific heat for different
 Rb$_{1-x}$Fe$_{2-y}$Se$_2$ samples.
 The arrow marks the anomaly in the specific heat at $T^*$=220~K for the sample BR~17.}
\end{figure}

In Fig.~14 the temperature dependences of the  specific heat is
shown in the representation $C/T$ vs. $T^2$ for different samples at
temperatures below 12~K. A linear dependence with two different
slopes below and above 5~K is clearly discernible. The values of the
residual Sommerfeld coefficient $\gamma_r$ were determined from the
fit to the experimental data at temperatures below 3~K using the
expression $C/T = \gamma_r + \beta T^2$, where the prefactor $\beta$
characterizes both the lattice and magnon contributions to the
specific heat, which are proportional to $T^3$, and therefore cannot
be separated. The calculated values of these parameters are given in
Table~\ref{tab2}. We notice a rather low value of $\gamma_r$ obtained for
the superconducting samples, which indicates their high purity and
minimal structural disorder or impurity content. The values of
$\gamma_r$ vary from 0.18 to 0.85~mJ/molK$^2$ for the samples with
high SC (BR~16) and poor SC parameters (F~266), respectively, reflecting
higher impurity content and disorder of the poorly SC samples in
agreement with the susceptibility results.
Similar values of $\gamma_r$ were reported for the related SC
K$_{1-\emph{x}}$Fe$_{2-\emph{y}}$Se$_2$ compound.\cite{BZ11} At the
same time, the non-superconducting samples BR~17 and BR~22 exhibit
rather high values of $\gamma_r$, which probably can be  attributed
to structural disorder due to Fe and Rb vacancies. Importantly, the
non-superconducting sample BR~19 exhibits an extremely low value of
$\gamma_r$ indicating minimal disorder, being correlated
with its best crystallinity revealed by x-ray studies (see
Fig.~2). It should be noted that the value of $\beta$ for the
sample BR~19 is the closest to those of the superconducting samples.
Together with very similar values of the susceptibility and its
temperature dependence (see Fig.~5), this justifies the use of the
specific heat data of the sample BR~19 as a reference for the
lattice and magnon contributions. The respective contributions for
the superconducting samples were calculated using the specific heat data of the
sample BR~19 scaled by the square root of the ratio of their molar mass.

\begin{figure}[htb]
\includegraphics[angle=0,width=0.45\textwidth]{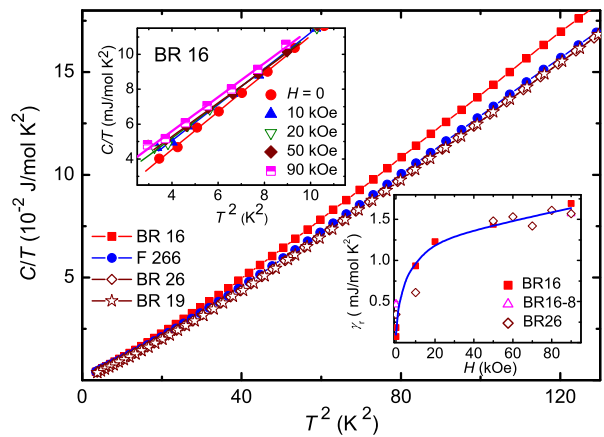}
 \caption{(color online) Temperature dependences of the specific heat in the
representation $C/T$ vs. $T^2$ for different Rb$_{1-x}$Fe$_{2-y}$Se$_2$ samples
at temperatures below 12~K. Upper inset: $C/T$ vs. $T^2$ in different applied
magnetic fields for the sample BR~16. Lower inset: magnetic field dependence
of the residual Sommerfeld coefficient $\gamma_r$ for different superconducting
samples with high SC parameters.
  }
\end{figure}

\begin{table*}[htb]
\caption{Parameters determined from the susceptibility ($T_c^{on}$, $\Delta T_c$) and heat capacity measurements.}\label{tab2}
\begin{tabular}{|l|c|c|c|c|c|c|}
\hline
Sample & $T_c^{on}$ & $\Delta T_c$ & $\gamma_r$ & $\beta$ & $\gamma_n$ &  $\Delta C$/$\gamma_n T_c$ \\
&  (K) &  (K)& (mJ/mol K$^2$) &  (mJ/mol K$^4$) & (mJ/mol K$^2$)  &\\
\hline F~266 & 30.9 & 2.8 & 0.85 & 0.96 & 10 & 1.0\\
\hline BR~16 & 32.4 & 1.5 & 0.18 & 1.09 & 12 & 0.95\\
\hline BR~16-8  & 32.4 & 1.6 & 0.46 &  1.04 &  12 & 0.96\\
\hline BR~17 &  -- & -- & 0.33 & 0.90 & --  & --\\
\hline BR~19 &  -- & -- & --  & 0.95  & -- & --\\
\hline BR~22 &  -- & -- & 3.0   & 1.36  & --  & --\\
\hline BR~26 & 32.4 & 1.5 &  0.34 &  0.94 & 7.4 &  2.2\\ \hline
\end{tabular}
\end{table*}

Fig. 15 shows the electronic specific heat in the representation $C_e/T$ vs. $T$
for several superconducting samples from different
batches obtained by subtraction of the lattice and magnon
contributions from the total measured specific heat. It has been
found that the superconducting contribution to the specific heat is
rather small compared to that of the lattice and magnon
contributions. However, a pronounced lambda-like anomaly of $C_{e}$
at the superconducting transition is clearly evidenced for all SC
samples. The sample F~266 with poor SC parameters exhibits a broad
anomaly of $C_{e}$ at $T_c$, which is displaced to lower temperatures
compared to the anomaly in samples from the batch BR~16 with high SC
parameters. The samples from batch BR~26 show the sharpest anomaly
of $C_{e}$ at $T_c$.

\begin{figure}[htb]
\includegraphics[angle=0,width=0.45\textwidth]{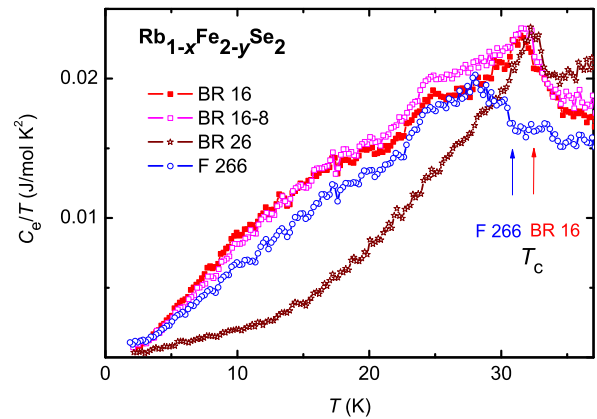}
 \caption{(color online) Temperature dependences of the electronic specific heat  $C_e/T$ vs. $T$
 for different Rb$_{1-x}$Fe$_{2-y}$Se$_2$ samples calculated by subtraction
 of the magnon and lattice contributions. The arrows mark $T_c$ for the samples F~266 and BR~16.}
\end{figure}

The value of the Sommerfeld coefficient  in the
normal state $\gamma_n$ for the  SC samples was estimated from the constraint of entropy
conservation at the onset of $T_c$, i.e., $\int_{0}^{T_c} C_e/T dT =
\int_{0}^{T_c} \gamma_n dT $.  The values of $\gamma_n$ are given in
Table II. They vary between 7.4 and 12 mJ/molK$^2$ for SC samples
from different batches and are comparable to those reported for SC
K$_{1-\emph{x}}$Fe$_{2-\emph{y}}$Se$_2$.\cite{BZ11} Assuming that
the residual Sommerfeld coefficient $\gamma_r$ corresponds to the
fraction of the normal conducting state, the obtained ratio of
$\gamma_r$/$\gamma_n$ implies that the volume fraction of the
superconducting phase in our samples varies within 92-98 \%. The
values of the jump of the specific heat $\Delta C/T_c$ at $T_c$ were
calculated by subtracting the value of $\gamma_n$ from the maximum
value of $C_e/T$ at $T_c$. They vary from 10 to 16 mJ/molK$^2$ for
different SC samples and are by a factor of $\sim5$ smaller than the
respective values of $\Delta C/T_c$ observed for single crystalline
FeSe$_{0.5}$Te$_{0.5}$.\cite{VT10,AS10,JHU11} The reduced specific
heat jump at $T_c$, $\Delta C$/$\gamma_n T_c$ varies from 0.95 (for
BR~16) to 2.25 (for BR~26). The values below 1.0 are probably related
to errors in the estimation of  $\gamma_n$ for these samples. The
obtained value of $\Delta C$/$\gamma_n T_c$ = 2.25 for the sample BR~26
with the sharpest anomaly in the specific heat at $T_c$ is higher
than the BCS estimate of 1.43 for the weak-coupling limit and
may indicate strong-coupling superconductivity in
Rb$_{1-x}$Fe$_{2-y}$Se$_2$ similar to that observed in Te substituted FeSe\cite{JHU11}
and in optimally doped  Ba$_{1-x}$K$_x$Fe$_2$As$_2$.\cite{GMU09,PP10,ChK10}

Since specific heat in the SC regime
varies substantially among samples with rather close stoichiometry,
we did not attempt to estimate the value of the superconducting gap.
However, to get an insight into the symmetry of the order parameter we measured the low-temperature specific heat  in
different magnetic fields. The upper inset in Fig. 14, where
the specific heat in the representation $C/T$ vs. $T^2$ is plotted
for one of the samples with high SC parameters (BR~16), demonstrates
the effect of the magnetic field on the specific heat at lowest
temperatures. The application of the magnetic field results in an
increase of the residual Sommerfeld coefficient $\gamma_r$ probably
related to the pair-breaking effect. The field dependences of
$\gamma_r$ for three samples with high SC parameters are shown in
the lower inset in Fig.~14. The values of $\gamma_r$ for these
samples are very close and exhibit a similar field dependence that
significantly differs from the linear dependence expected for fully
gapped superconductors.  It must be noted that the dependence of $\gamma_r(H)$ is
close to $\sqrt{H}$ characteristic for \emph{d}-wave symmetry.\cite{GEV93}
Recently, a Volovik-like $\sqrt{H}$ term in the specific heat
was reported for an optimally doped BaFe$_2$(As$_0.7$P$_0.3)_2$ pnictide which indicate
the presence of nodes in the superconducting gap.\cite{YWa11}
However, there is no experimental evidence of nodes in
A$_{1-\emph{x}}$Fe$_{2-\emph{y}}$Se$_2$ yet. On the contrary, very
recent ARPES studies of related A$_{1-\emph{x}}$Fe$_{2-\emph{y}}$Se$_2$ materials\cite{DM11, LZ11}
seem to fully rule out $d$-wave pairing. At present, the sub-linear dependence of the $\gamma_r(H)$ observed in Rb$_{1-\emph{x}}$Fe$_{2-\emph{y}}$Se$_2$ is
lacking any reasonable theoretical explanation but multiband effect with two different gaps discussed in Ref.~\onlinecite{YWa11} can be also relevant in this case.

\section{Summary and Conclusions}

Our detailed preparative, structural, magnetic, conductivity, and
specific heat studies of Rb$_{1-x}$Fe$_{2-y}$Se$_2$ single crystals
testify several important peculiarities of these materials:

1. Three different regions with distinct structural, magnetic, and
conducting behavior of the samples are documented in the
Rb$_{1-x}$Fe$_{2-y}$Se$_2$ system as presented in the phase diagram
in Fig. 16. Superconductivity is observed in the range of the Fe
concentration above 1.53 and below 1.60 (region II of the diagram).
Superconducting samples with the sharpest transitions, the highest
critical temperature of 32.4 K and the largest diamagnetic response
have the "optimal" composition Rb$_{0.8}$Fe$_{1.6}$Se$_2$.
The reduction of the superconducting parameters of the samples can be attributed
to deviations of the Fe concentration from the "optimal" value of 1.6.
Within the experimentally determined range of the Rb concentration (0.6--0.8) no correlation
between the variation of the Rb content and the lattice parameters of the samples was found.
The superconducting samples show the smallest value of the lattice parameter \emph{c}
compared to the non-superconducting samples.
Samples with the Fe content below 1.5 (region I) and above 1.6
(region III) show an insulating, or semiconducting behavior at low
temperatures, respectively.

\begin{figure}[htb]
\includegraphics[angle=0,width=0.5\textwidth]{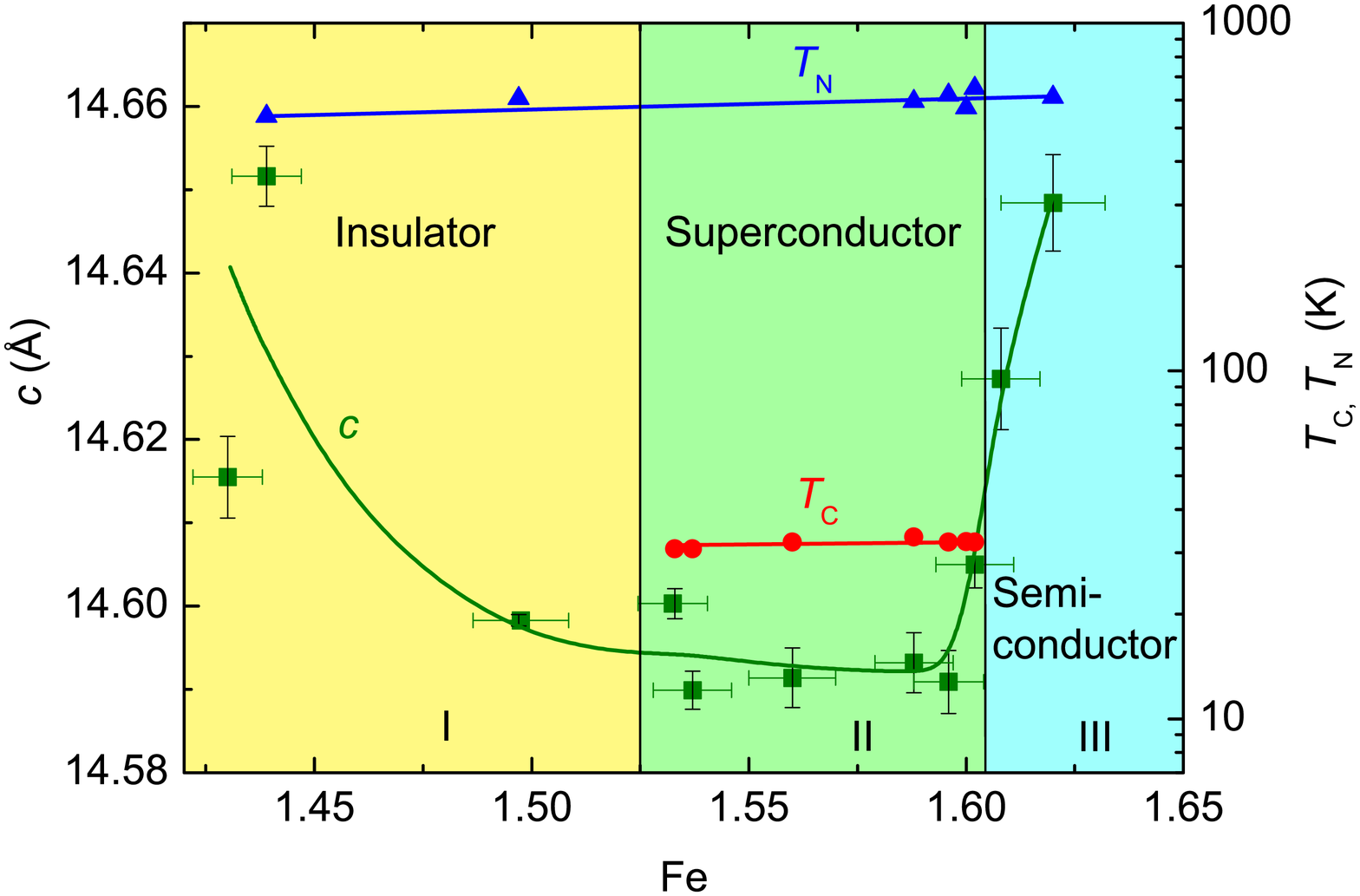}
\caption{(color online) Phase diagram of Rb$_{1-\emph{x}}$Fe$_{2-\emph{y}}$Se$_2$
system presenting dependences of the lattice constant \emph{c}, transition temperature $T_c$ and
the N\'{e}el temperature $T_N$  for three different regions with distinct structural,
magnetic and conducting behavior. }
\end{figure}

2. In the whole range of the studied Fe concentrations the
susceptibility measurements evidenced a behavior typical for an
anisotropic antiferromagnet with a transition temperature $T_N$  above 500~K.

3. Magnetization measurements provide an evidence for the  presence
of antiferromagnetic correlations in the ordered superconducting
state in samples with the Fe concentration in the range of 1.53--1.6.
This suggests the coexistence of superconductivity and static
antiferromagnetic order. In a view of recent experiments \cite{JTP11,ACh11,VKs11} this coexistence should be considered
as being on the macroscopic level within a scenario of the phase separation.

4. Magnetic hysteresis measurements of the superconducting samples
revealed high values of the
critical current density $j_c$ of $1.6 \times 10^4$~A/cm$^2$ for
superconducting samples with the composition close to
Rb${_2}$Fe${_4}$Se${_5}$. The obtained value of $j_c$  is the
largest reported thus far for A$_{1-\emph{x}}$Fe$_{2-\emph{y}}$Se$_2$ materials.

5. The upper critical fields $H_{c2}$ of $\sim 250$~kOe for the
in-plane, and 633~kOe for the inter-plane configurations are
estimated from the resistivity study of the superconducting samples
in magnetic fields. They are the
lowest reported to date for the Rb$_{1-x}$Fe$_{2-y}$Se$_2$
superconductors.

6. Specific-heat measurements of the superconducting samples evidenced
very low values of the residual Sommerfeld coefficient corresponding
to 92--98\% volume fraction of the superconducting phase, which confirm
the bulk nature of the superconducting state.

7. Resistivity  studies revealed an insulating behavior at low
temperatures for samples with Fe concentrations below 1.5. The
conductivity of these samples at low temperatures can be reasonably
described by the variable-range hopping Mott mechanism. The
conductivity of the samples with Fe concentrations above 1.6 is
typical for heavily doped semiconductors.

8. In the non-superconducting samples with the Fe concentration below 1.45
an unusual phase transition below 233~K is revealed which is not related to
antiferromagnetic or structural transformations.

9. The discrepancies in the range of Fe concentrations, where
superconductivity shows up in Rb$_{1-x}$Fe$_{2-y}$Se$_2$ as compared
to earlier reports, are likely to be related to the accuracy of the
methods used for the compositional analysis.

\begin{acknowledgements}

The authors thank Dana Vieweg for experimental support. This
research has been supported by the DFG via SPP 1458 under Grant DE1762/1-1 and the
Transregional Collaborative Research Center TRR 80 (Augsburg - Munich).

\end{acknowledgements}

\end{document}